\documentclass[a4paper,13pt]{article}

\usepackage{times}                       
\usepackage{CJK,CJKnumb,CJKulem}         
\usepackage{color}                       
\usepackage[all]{xy}
\usepackage{amsmath,amsthm,amsfonts,amssymb,bm} 
\usepackage{graphicx,psfrag}                    
\usepackage[all]{xy}
\usepackage{float}
\newtheorem{prop}{Proposition}[section]
\newtheorem{defi}{Definition}[section]
\newtheorem{exam}{Example}[section]
\newtheorem{lemm}{Lemma}[section]

\begin{document} 

\author{Qizhi Zhang, Deping Liu}                                 
\title{On the hopping pattern design for D2D Discovery}                                

\section{Introduction}

D2D ( device to device ) UEs discover each other through D2D discovery procedure. Due to half duplex mode,
D2D UEs cannot receive discovery signals from other D2D UEs while transmitting its own; and a closer UE¡¯s discovery
signal may severely interfere with a farther one¡¯s, even if they use different sub band, because of in-band emission. Therefore, a D2D UE¡¯s transmit and receive
 chances should be carefully arranged to enable the UE to discover as many UEs as possible. A hopping discovery pattern is
 designed to meet this requirement.

 In \cite{QC}, a periodic discovery frame is introduced for this purpose. A discovery frame is
divided into $n$ sub frame. In addition, the frequency band of system is split into $m$
 parallel channels using OFDMA, where $m\leq n$. A frequency-time
unit shown in Figure 1 corresponds to a unique discovery resource unit and is the basic resource unit for the
device to send or receive a discovery signal. Hence, in one discovery cycle, there are $mn$ discovery resource units and each
device selects one resource unit to transmit its discovery signal according to the following hopping pattern:
\begin{align*}
& i(t)=i(0) \\
& j(t)=(j(0)+i(0)t) \mod n.
\end{align*}
\begin{figure}[H]
\centering
\includegraphics[width=\textwidth]{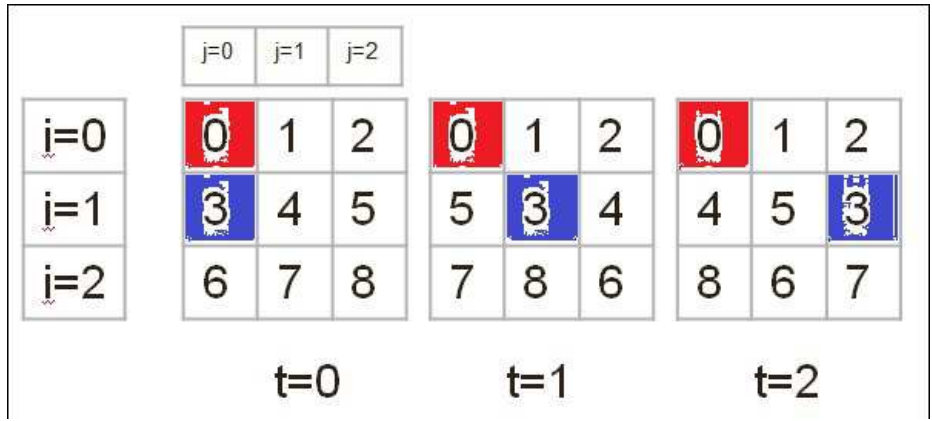}
\caption{Example of hopping pattern}
\end{figure}

The two equations define a UE's transmission frequency-time unit $(i(t), j(t))$ in discovery frame $t$,
 which is decided by the frequency time unit $(i(0), j(0))$  in discovery frame $0$. For example,
 if UE 0 transmits its discovery signal on red units ,witch $(i(0), j(0))=(0,0), (i(1), j(1))=(0, 0), \cdots$,
and UE 3 transmits its discovery signal on blue units, witch $(i(0), j(0))=(1,0), (i(1), j(1))=(1,1), \cdots$,
though they can not receive each other?¡¥s discovery signals in discovery frame $0$, yet in frame $1$ they can receive the signals.

Once hopping pattern is determined, some relation is created between the discovery resource units in differential discovery frames. For example, we can see the red frequency-time units as one logical discovery resource, and the blue frequency-time units as another logical discovery resource.

\section{Formulation}

A discovery frame is
divided into multiple sub frames. Let $J$ denote the set of these sub frames. The system frequency band is split into some
 parallel channels using OFDMA. Let $I$
denote the set of these parallel channels.

Then the resources in every discovery frame can be divided into $|I| \times |J|$ units, called frequency-time resource,
each of which corresponds to an element in set $I \times J$. This discovery frame structure can allow $|I| \times |J|$ UE¡¯s
to transmit their discovery signals, each UE selects a different frequency-time resource. The frequency-time resource used by a
UE should change in every discovery frame so as to hear as many as UEs due to half duplex and to randomize the
interference caused by in-band emission.

Let $S$ be the a set  of $|I|\times|J|$ elements (logical discovery resources). We call a sequence of mapping
\begin{align}
	(i(t),j(t)):S \longrightarrow I \times J  \quad \mbox{ for any } t\in \mathbb{Z}	
\end{align}
a frequency-time hopping pattern, or a hopping pattern on discovery frame structure $I\times J$, denote it by
$\{ (i(t),j(t)) \}_{t \in \mathbb{Z}}$. For a logical discovery resource $s \in S$, call $i(t)(s), j(t)(s), (i(t)(s),j(t)(s))$ the frequency coordinate,
 the time coordinate, the frequency time coordinate of $s$ in discovery frame $t$ respectively.

Every UE selects a logical discovery resource $s$ (different UEs select different logical resources)
, and transmit own discovery signal
on frequency resource $(i(t)(s),j(t)(s))$ in every discovery frame $t$. Two example of hopping pattern are given below:

\begin{exam}
 Let $\{ (i(t)(s), j(t)(s))\}_{t \in \mathbb{Z}, s\in S}$ be i.i.d random variable uniform distributed on $I \times J$, then we call
 the hopping pattern $\{ (i(t), j(t)\}_{t \in \mathbb{Z}}$ random pattern on discovery frame structure $I \times J$.
\end{exam}

\begin{exam}
Let $(i(0),j(0)): S \longrightarrow \mathbb{Z}/m\mathbb{Z} \times \mathbb{Z}/n\mathbb{Z}$ be any bijection, and define
$\{(i(t),j(t))\}_{t \in \mathbb{Z}}$ by
\begin{align*}
i(t)&=(i(0)+kt) \mod m    \\
j(t)&=(j(0)+i(0)t+\mbox{floor}(i(0)/n)t^2) \mod n
\end{align*}
where $k$ is a constant integral number. This pattern is known as QC¡¯s hopping pattern on discovery frame
structure $I \times J=\mathbb{Z}/m\mathbb{Z} \times \mathbb{Z}/n\mathbb{Z}$ (\cite{R1}). It is easy to see that when $m \leq n$, let $k=0$, then this
pattern degenerate into the pattern in section 1.
\end{exam}

Let $\{(i(t), j(t))\}_{t \in \mathbb{Z}_{\geq 0}}$ be a hopping pattern. If there exists a positive integral
 number $T$ satisfying the following condition (P), we call the hopping pattern column periodical.

(P): for any $s, s' \in S, t\in \mathbb{Z}$,
\begin{align*}
j(t)(s)=j(t)(s') \mbox{ if and only if } j(t+T)(s)=j(t+T)(s').
\end{align*}

If the hopping pattern is column periodical, we call the hopping pattern {\bf has column period $T$}, where $T$ is the minimal positive
 integral number satisfying the condition (P); otherwise we call the hopping pattern {\bf has column period infinity}.

 Suppose the
column period is $T$ and fast fading is not considered, then any UE can not discover new UE after $T$ discovery frames.
Even
we consider the channel stage variety brought by fast fading, from $T$th discovery frame, the newly discovered UEs are
 relatively few. As a result, a hopping pattern having a small column period is not a good choice.

If for any $s, s' \in S$, there exists a real number $\rho _{s,s'}$ such that
\begin{align*}
\frac{1}{t_b}
\sharp \{t=0, 1, \cdots, t_b-1 \mid  j(t)(s)=j(t)(s') \}
\end{align*}
converges in probability to $\rho _{s,s'}$, we call
\begin{align*}
\begin{array}{c}
\mbox{max} \\
s, s'\in S, s\neq s'
\end{array} \rho_{s,s'}
\end{align*}
the {\bf maximal collision ratio} of this pattern.

The maximal collision ratio measures the fairness of hopping pattern. For example, if the maximal collision ratio of some hopping
pattern is $1$, it means that there are two logical resources $s$ and $s'$ such, that the UE $A$ using logical resource $s$ and
the UE $A'$ using logical resource $s'$ almost always transmit discovery signals at the same time, hence they are almost
never discoverable
for each other. But other UE pair may be have a lot of chance to discover each other. It is unfair for this UE pair $(A, A')$.

We define the {\bf maximal continual collision number} as
\begin{align*}
\mbox{max} \{l=1, 2, ... \mid  &  \mbox{ There exist } s, s', t \mbox{ such that } j(t)(s)=j(t)(s'), \\
& j(t+1)(s)=j(t+1)(s'), \cdots, j(t+l-1)(s)=j(t+l-1)(s')\},
\end{align*}
whose value is either a positive integral number or $\infty$;

\begin{prop}
The maximal continual collision number of any hopping pattern on discovery frame structure $\mathbb{Z}/m\mathbb{Z} \times \mathbb{Z}/n\mathbb{Z}$ is greater than or equal to
 $\log _n(m)$. 
\end{prop}
{\bf Proof.} Suppose the maximal continual collision number is less than $\log _n(m)$. Let $r$ be the minimal positive integer that greater than or
equal to $\log _m(n)$, then there are no $s\neq s' \in S$ such, that $j(t)(s)=j(t)(s')$, for $t=0, 1, \cdots r-1$. Therefore the map
\begin{align*}
\begin{array}{ccl}
 S & \longrightarrow & \mathbb{Z}/n\mathbb{Z} \times \mathbb{Z}/n\mathbb{Z} \times \cdots \mathbb{Z}/n\mathbb{Z} \\
 s & \mapsto & (j(0)(s), j(1)(s), \cdots j(r-1)(s))
\end{array}
\end{align*}
is an injection. Therefore the number of elements in $S$ is less than $n^r$, i.e $mn<n^r$. On the one hand,
we know $r<log_n(m)+1$, hence $mn<n^{log_n(m)+1}=mn$, which is contradictory.
\qed

\begin{defi}
Call a hopping pattern {\bf local good pattern} if its maximal continual collision number is the minimal integral number greater than or equal to $\log_n(m)$.
\end{defi}

\begin{exam}
Let $\{(i(t), j(t))\}_{t \in \mathbb{Z}_{\geq 0}}$ be a random hopping pattern on discovery frame structure $\mathbb{Z}/m\mathbb{Z} \times \mathbb{Z}/n\mathbb{Z}$. Then the column period of this pattern is $\infty$.
  For any $s \neq s' \in S$, we know that $\{ j(t)(s)-j(t)(s')\}_{t \in \mathbb{Z}_{\geq 0}}$ are independent uniform random variables
  taking value in $\mathbb{Z}/n\mathbb{Z}$, hence by law of large numbers we know that
\begin{align*}
\frac{1}{t_b} \sharp \{t=0, 1, \cdots, t_b-1 \mid  j(t)(s)-j(t)(s')=0 \}
\end{align*}
converges in probability to $\frac{1}{n}$, i.e its maximal collision ratio is $\frac{1}{n}$.
It is easy to see, its maximal continual collision number is $\infty$, hence it is not a local good pattern.
\qed
\end{exam}

\begin{exam}
Let $n=p$ be an odd prime number and $m$ be a positive integral number. Let
$\{(i(t), j(t))\}_{t \in \mathbb{Z}_{\geq 0}}$ be the QC's pattern on discovery frame structure $\mathbb{Z}/m\mathbb{Z}
\times \mathbb{Z}/p\mathbb{Z}$.
 It is easy to see that its column period is $p$.

When $m\leq p$, the QC pattern degenerates to
\begin{align*}
i(t)&=(i(0)+kt) \mod m    \\
j(t)&=(j(0)+i(0)t) \mod p
\end{align*}
Let $s \neq s'$ be any two different resources, then we know $(i(0)(s), j(0)(s)) \neq (i(0)(s'), j(0)(s'))$.
Because the polynomial equation about $t$ of degree at most one:
\begin{align*}
j(0)(s)+i(0)(s)t =j(0)(s')+i(0)(s')t
\end{align*}
 has at most one root in the finite field $\mathbb{F}_p$, we know that the maximal collision times of $s$ and $s'$ is not
 greater than $1$ in a column period. Therefore its maximal collision ratio is $\frac{1}{p}$, its maximal
 continual collision number is $1$.

 When $p<m \leq p^2$, let $s \neq s'$ be any two different resources, then we know $(i(0)(s), j(0)(s)) \neq (i(0)(s'), j(0)(s'))$. Because the polynomial equation about $t$ of degree at most two:
\begin{align*}
j(0)(s)+\mod(i(0)(s), p)t+\mbox{floor}(i(0)(s)/p)t^2 = \\
j(0)(s')+\mod(i(0)(s'), p)t+\mbox{floor}(i(0)(s')/p)t^2
\end{align*}
 has at most two roots in the finite field $\mathbb{F}_p$, we know that the maximal collision times of $s$ and $s'$ is not
 greater than $2$. Now consider the two special resources $s$ and $s'$ defined by
\begin{align*}
(i(0)(s), j(0)(s))=(1, 0), \quad (i(0)(s'), j(0)(s'))=(n,0).
\end{align*}
We know that
\begin{align*}
 j(t)(s)=t \mod n, \quad j(t)(s')=t^2 \mod n,
\end{align*}
hence $j(t)(s)=j(t)(s')$ for all $t \equiv 0, \mbox{ or } 1 \mod n$. Therefore the maximal collision ration of QC's hopping
pattern is $\frac{2}{p}$, its continual collision number is $2$.

When $m>p^2$,  consider the two special resources $s$ and $s'$ defined by
\begin{align*}
(i(0)(s), j(0)(s))=(0, 0), \quad (i(0)(s'), j(0)(s'))=(p^2,0).
\end{align*}
Because
\begin{align*}
\mod(i(0)(s), p)=&\mod(i(0)(s'), p), \\
\mbox{floor}(i(0)(s)/p)=&\mbox{floor}(i(0)(s')/p),
\end{align*}
they are will collide constantlly. Therefore the maximal collision ratio is $1$, the maximal continual collision number is $\infty$.

In summary, when $m \leq p^2$ this pattern is a local good pattern, its maximal continual collision
number is ${ceil}(\log_p m)$; when $m>p^2$ the maximal collision ratio is $1$, the maximal continual collision
number is $\infty$. \qed
\end{exam}

If we compare random pattern and QC pattern, we see that the random pattern has large column period $(\infty)$
and lower maximal collision ratio, but has higher maximal continual collision number. The QC pattern has  acceptable
 column period, maximal collision ratio and maximal
 collision number if $n \geq \sqrt{m}$ is an odd prime number, but its performance degrades considerably with respect
 to the three metrics when $n<\sqrt{m}$.

In general $m$ is determined by entire system bandwidth and is constant. But the number of UEs may change, hence $n$ may change. We wish the three metrics still
good even when $n$ is little.

\section{A new pattern}

Now, we construct a class of local good hopping patterns whose
 column period does not shorten as $J$ decreases and at the same time whose maximal collision ratio is small.

Let $n=p$ be a prime number, $r$ be the minimal integral number greater than or equal to $log_p(m)$. Let $c_0, c_1, \cdots, c_{r-1}: \mathbb{Z}/m\mathbb{Z} \longrightarrow \mathbb{Z}/p\mathbb{Z}$ be the $p$-adic representation coefficient maps, i.e. for any $i\in \mathbb{Z}/m\mathbb{Z}$, $i$ can be represented uniquely as
\begin{align*}
i=c_0(i)+c_1(i)p+\cdots c_{r-1}(i)p^{r-1}
\end{align*}
where $c_0(i), c_1(i), \cdots, c_{r-1}(i) \in \mathbb{Z}/p\mathbb{Z}$.

Let $f(x)=x^r+a_1x^{r-1}+ \cdots +a_r$ be an irreducible polynomial on $\mathbb{F}_p$ satisfying the following condition (G):

(G): The minimal integral number $a$ satisfying
\begin{align*}
x^{a} \equiv 1 \mod f(x) \mbox{     ( as the polynomials in $\mathbb{F}_p[x]$ )}
\end{align*}
is $p^r-1$. ( In fact, the image of such $x$ under the ring homomorphism  $\mathbb{F}_p[x] \longrightarrow \frac{\mathbb{F}_p[x]}{(f(x))} \backsimeq \mathbb{F}_{p^r}$ is a generator of the multiplicative group $\mathbb{F}_{p^r}^\times$. )
\qed

Let $(i(0),j(0)): S \longrightarrow \mathbb{Z}/m\mathbb{Z} \times \mathbb{Z}/p\mathbb{Z}$ be any bijection,
and define $\{(i(t), j(t))\}_{t \in \mathbb{Z}_{> 0}}$ as follows:
\begin{align*}
\begin{array}{ccll}
i(t)& =& i(0)+kt & \mod m  \\
j(t)& = & j(0)+(c_0(i(0)), c_1(i(0)), \cdots , c_{r-1}(i(0)))b(t) & \mod p
\end{array}
\end{align*}
where $k$ is a constant integral number, for any integral number $t$, $b(t)$ is a column
vector in $\mathbb{F}_p^{r}$ defined as
\begin{align*}
b(t)=\left\{
\begin{array}{cl}
0 & \mbox{ if } t \equiv 0 \mod p^r  \\
A^{(t \mod p^r)-1}b  & \mbox{ else }
\end{array}
\right.
\end{align*}
where $A$ is a matrix on $\mathbb{F}_p$ defined by
\begin{align*}
A=\left\{
\begin{array}{cc}
\left(
\begin{array}{ccccc}
0 & 0 & \cdots & 0 & -a_r \\
1 & 0 & & \vdots & -a_{r-1} \\
0 & 1 & \ddots & \vdots \\
\vdots & \ddots & \ddots & 0 & \vdots \\
0 & \cdots & 0 & 1 & -a_1
\end{array}
\right) & \mbox{ if } r>1 ; \\
-a_1  &  \mbox{ if } r=1
\end{array}
\right.
\end{align*}
and $b$ is any constant column vector in $\mathbb{F}_p^r \setminus \{0\}$.

\begin{prop}
The hopping pattern $\{(i(t), j(t))\}_{t \in \mathbb{Z}_{\leq 0}}$ defined previously is a local good pattern, its column period is
$p^r$, its maximal collision ratio is $\frac{1}{p}$.
\end{prop}

{\bf Proof.}

{\bf Local.} Suppose there are two different resources $s$ and $s'$, and an integral number $t$ such that
\begin{align*}
\begin{array}{rcl}
j(t)(s)&=&j(t)(s') \\
j(t+1)(s)&=&j(t+1)(s') \\
&\vdots & \\
j(t+r)(s)&=&j(t+r)(s').
\end{array}
\end{align*}
Denote $(c_0(i(0)(s)), c_1(i(0)(s)), \cdots , c_{r-1}(i(0)(s)))$ by $\alpha (s)$, $(c_0(i(0)(s')), c_1(i(0)(s')), \cdots , c_{r-1}(i(0)(s')))$ by $\alpha (s')$, then we have
\begin{align*}
(j(0)(s), \alpha (s)) \left(
\begin{array}{cccc}
1 & 1 & \cdots & 1 \\
b(t) & b(t+1) & \cdots & b(t+r)
\end{array}
\right)=(j(0)(s'), \alpha (s')) \left(
\begin{array}{cccc}
1 & 1 & \cdots & 1 \\
b(t) & b(t+1) & \cdots & b(t+r)
\end{array}
\right)
\end{align*}
as the elements in $\mathbb{F}_p^{r+1}$. But we know the matrix
\begin{align*}
\left(
\begin{array}{cccc}
1 & 1 & \cdots & 1 \\
b(t) & b(t+1) & \cdots & b(t+r)
\end{array}
\right)
\end{align*}
is non-singular from the following lemma \ref{non-singular}. Therefore we have
\begin{align*}
(j(0)(s), \alpha (s))=(j(0)(s'), \alpha (s')).
\end{align*}
It contradicts that $(i(0), j(0)): S \longrightarrow \mathbb{Z}/m\mathbb{Z} \times \mathbb{Z}/n\mathbb{Z}$ is bijective. Therefore $\{(i(t), j(t))\}_{t \in \mathbb{Z}_{\leq 0}}$ is a local good pattern.

 {\bf Period.} It easy to see, $p^r$ satisfies the condition (P). we just need to prove that $p^r$ is the minimal positive integral
number satisfying the condition (P). Suppose there exists a positive integral number $T<p^r$ satisfies the condition (P).

Let $s_0$ be the logical discovery resource such that $j(0)(s_0)=0$ and the coefficients of the p-adic representation of
$i(0)(s_0)$ is $(0, 0, \cdots, 0, 0)$.

Let $s_1$ be the logical discovery resource such that $j(0)(s_1)=0$ and the coefficients of  the p-adic representation of
$i(0)(s_1)$ is $e_1:=(1, 0, \cdots, 0, 0)$.

Let $s_2$ be the logical discovery resource such that $j(0)(s_2)=0$ and the coefficients of  the p-adic representation of
$i(0)(s_2)$ is $e_2:=(0, 1, 0, \cdots, 0)$.

$\cdots$

Let $s_r$ be the logical discovery resource such that $j(0)(s_r)=0$ and the coefficients of  the p-adic representation of
$i(0)(s_r)$ is $e_r:=(0, 0, \cdots, 0, 1)$.

Because $j(0)(s_i)=j(0)(s_0)$ for all $i=1, 2, \cdots, n$, we have
\begin{align*}
j(T)(s_i)=j(T)(s_0)       \mbox{ for all } i= 1, 2, \cdots, n.
\end{align*}
Therefore
\begin{align*}
r_ib(T)=0 \mbox{ for all } i=1, 2, \cdots, n.
\end{align*}
Hence $b(T)=0$. But in a period $b(t)$ take every value in $\mathbb{F}_p^{r}$ once by following lemma
(\ref{once}), which is contradictory to that $b(0)=b(T)=0$.

{\bf Collision ratio.} Suppose two different logical discovery resources $s$ and $s'$ have a collision at discovery frame $t$, i.e
\begin{align*}
j(t)(s)-j(t)(s')=0
\end{align*}
Denote $(c_0(i(0)(s)), c_1(i(0)(s)), \cdots , c_{r-1}(i(0)(s)))$ by $\alpha (s)$, and denote $(c_0(i(0)(s')), c_1(i(0)(s')), \cdots , c_{r-1}(i(0)(s')))$ by $\alpha (s')$, then we have
\begin{align*}
(\alpha(s)-\alpha(s'))b(t)=-(j(0)(s)-j(0)(s')).
\end{align*}
Note that the period of $b(t)$ is $p^r$, in a period $b(t)$ take every value in $\mathbb{F}_p^{r}$ once by following lemma
(\ref{once}), hence the number of solutions $\beta \in \mathbb{F}_p^{r}$ of equation
\begin{align}
\label{rank1}
(\alpha(s)-\alpha(s'))\beta=-(j(0)(s)-j(0)(s'))
\end{align}
is equal to the number of collisions in a period $p^r$ of $b(t)$. When $\alpha(s)=\alpha(s')$, because
 $(i(0),j(0)): S \longrightarrow \mathbb{Z}/m\mathbb{Z} \times \mathbb{Z}/n\mathbb{Z}$ is a bijection,
  we know that $j(0)(s)\neq j(0)(s')$, hence the equation (\ref{rank1}) has no solution, i.e. the resources
  $s$ and $s'$ have no collision.
When $\alpha(s) \neq \alpha(s')$, the rank of both coefficient matrix and augmented matrix is $1$,
 hence the set of solutions of equation (\ref{rank1}) is a coset of a linear subspace of dimension
  $r-1$ in $\mathbb{F}_p^r$, i.e. the number of solutions of the equation (\ref{rank1}) is $p^{r-1}$.
   Therefore the maximal collision ration is
\begin{align*}
\frac{p^{r-1}}{p^r}=\frac{1}{p}.
\end{align*}          \qed

\begin{lemm}
\label{non-singular}
\begin{align}
\label{non_singular_matrix}
\left(
\begin{array}{cccc}
1 & 1 & \cdots & 1 \\
b(t) & b(t+1) & \cdots & b(t+r)
\end{array}
\right)
\end{align}
is a non-singular matrix in $\mathbf{M}_{r+1}(\mathbb{F}_p)$.
\end{lemm}

{\bf Proof.}
Note that the period of $b(t)$ is $p^r$. Therefore:

 {\bf (a).} If there is not a number in $t, t+1, \cdots , t+r$  can be divided exactly by $p^r$, we can suppose $1 \leq t$ and $t+r \leq p^r-1$. To prove
\begin{align*}
\left(
\begin{array}{cccc}
1 & 1 & \cdots & 1 \\
b(t) & b(t+1) & \cdots & b(t+r)
\end{array}
\right)
\end{align*} is  non-singular we just need to proof that the matrix
\begin{align*}
\left(
\begin{array}{ccccc}
1 & 0 & 0 &  \cdots & 0 \\
b(t) & b(t+1)-b(t) & b(t+2)-b(t+1) & \cdots & b(t+r)-b(t+r-1)
\end{array}
\right)
\end{align*}
is non-singular, hence just need to prove that the column vectors
\begin{align*}
\begin{array}{cccc}
 b(t+1)-b(t) ,& b(t+2)-b(t+1), & \cdots , & b(t+r)-b(t+r-1)
\end{array}
\end{align*}
are linear independent on $\mathbb{F}_p$. In fact, they are
\begin{align*}
A^{t-1}(A-I)b, A^{t}(A-I)b, \cdots A^{t+r-2}(A-I)b
\end{align*}
Because the character polynomial of $A$ on $\mathbb{F}_p$ is $f(x)$, it is irreducible, we know $A$ and $A-I$ are non-singular. Hence we just need to show that
\begin{align*}
b, Ab, \cdots , A^{r-1}b
\end{align*}
are linear independent on $\mathbb{F}_p$. Suppose they are linear dependent on $\mathbb{F}_p$, i.e there exists
 $d_0, d_1, \cdots d_{r-1}$, not all zero, such that
\begin{align*}
d_0b+d_1Ab+ \cdots d_{r-1}A^{r-1}b=0 \in \mathbb{F}_p^{r}.
\end{align*}
Let $g(x)=d_0+d_1x+\cdots +d_{r-1}x^{r-1} \in \mathbb{F}_p[x]$, then we have
\begin{align*}
g(A)b=0 \in \mathbb{F}_p^r
\end{align*}
On the other hand, we have
\begin{align*}
f(A)b=0 \in \mathbb{F}_p^r
\end{align*}
also, hence $h(A)b=0 \in \mathbb{F}_p^r$, where $h(x)$ is the greatest common divisor of $f(x)$ and $g(x)$.
But $f(x)$ is irreducible, hence $h(x)=1$, hence $b=0$. It is contradictory to that $b \in \mathbb{F}_p ^r \setminus \{0\}$.

 {\bf (b).} If there is a number in $t, t+1, \cdots , t+r$  can be divided exactly by $p^r$, for example, $t+e$ is divided by $p^r$, the matrix (\ref{non_singular_matrix}) equal to
\begin{align*}
\left(
\begin{array}{ccccccc}
1 &  \cdots &1& 1&1 & \cdots & 1\\
b(-e) &  \cdots &b(-1) & b(0)& b(1) & \cdots & b(r-e)
\end{array}
\right)
\end{align*}
We know $b(0)=0$, hence we just need to show that
\begin{align*}
b(-e) ,  \cdots , b(-1) , b(1) , \cdots , b(r-e)
\end{align*}
are linear independent on $\mathbb{F}_p$, i.e
\begin{align*}
A^{p^r-e-1}b, \cdots, A^{p^r-2}b, A^0b, \cdots, A^{r-e-1}b
\end{align*}
are linear independent on $\mathbb{F}_p$. Note that $A^{p^r-1}=I$ because $x^{p^r-1}-1 \in \mathbb{F}_p[x]$ is divided by the character polynomial $f(x)$ of $A$. Therefore we just need to show that
\begin{align*}
A^{p^r-e-1}b, \cdots, A^{p^r-2}b, A^{p^r-1}b, \cdots, A^{p^r+r-e-2}b
\end{align*}
are linear independent $\mathbb{F}_p$, that is like in the part (a).
    \qed

\begin{lemm}
\label{once}
In a period $p^r$ , $b(t)$ takes every value in $\mathbb{F}_p^r$ once.
\end{lemm}
{\bf Proof.}

We just need to prove that the map
\begin{align*}
\begin{array}{rcl}
\mathbb{Z}/(p^r-1)\mathbb{Z} & \longrightarrow &\mathbb{F}_p^r \setminus \{0\} \\
t & \mapsto & A^tb
\end{array}
\end{align*}
is bijective.

Suppose there are $t_1<t_2 \in \mathbb{Z}/(p^r-1)\mathbb{Z}$ such that $A^{t_1}b=A^{t_2}b$. Let $u(x)=x^{t_2-t_1}-1 \in \mathbb{F}_p[x]$, then we have
\begin{align*}
u(A)b=0.
\end{align*}
On the other hand, we know $f(A)b=0$. Therefore we have
\begin{align*}
v(A)b=0,
\end{align*}
 where $v(x)\in \mathbb{F}_p[x]$ is the greatest common divisor of $f(x)$ and $u(x)$.
But $f(x)$ is irreducible, hence $v(x)=1$ or $f(x)$. But we know $b \neq 0$, hence $v(x) \neq 1$. Therefore $v(x)=f(x)$,
 hence $u(x)$ is exactly divided by $f(x)$. This is a contradiction to that $p^r-1$ is the minimal positive integral number $a$ such that $x^a \equiv 1 \mod f(x)$. Therefore the assumption is false, this map is an injection. On the other hand,
the numbers of elements in  $\mathbb{Z}/(p^r-1)\mathbb{Z}$ and $\mathbb{F}_p^r \setminus \{0\}$ are same, hence this map is a bijection.
 \qed

\section{Polynomials satisfying condition (G)}
We list some polynomials satisfying the condition (G) when $p$ is a prime number and less than $50$:

\begin{tabular}{|c|c|c|c|}
\hline
m & p & r & f(x) \\
\hline
33 $\sim$ 64 & 2 & 6 & $x^6+x^5+x^3+x^2+1$\\
\hline
4 $\sim$ 9   & 3 & 2 & $x^2-x-1$\\
\hline
10 $\sim$ 27 & 3 & 3 & $x^3+2x^2+x+1$\\
\hline
28 $\sim$ 81 & 3 & 4 & $x^4+2x+2$ \\
\hline
26 $\sim$ 125 & 5 & 3 & $x^3+4x^2+x+2$\\
\hline
50 $\sim$ 343 & 7 & 3 & $x^3+5x+2$\\
\hline
8 $\sim$ 49 & 7 & 2 & $x^2+6x+3$\\
\hline
12 $\sim$ 121 & 11 & 2 & $x^2+3x+6$\\
\hline
14 $\sim$ 169 & 13 & 2 & $x^2+4x+2$\\
\hline
18 $\sim$ 289 & 17 & 2 & $x^2+15x+12$\\
\hline
20 $\sim$ 361 & 19 & 2 & $x^2+12x+2$\\
\hline
24 $\sim$ 529 & 23 & 2 & $x^2+10x+10$\\
\hline
30 $\sim$ 841 & 29 & 2 & $x^2+22x+19$\\
\hline
32 $\sim$ 961 & 31 & 2 & $x^2+16x+3$\\
\hline
38 $\sim$ 1369 &  37 & 2 & $x^2+12x+19$\\
\hline
42 $\sim$ 1681 &  41 & 2 & $x^2+9x+29$\\
\hline
44 $\sim$ 1849 &  43 & 2 & $x^2+25x+26$\\
\hline
48 $\sim$ 2304 &  47 & 2 & $x^2+14x+10$\\
 \hline
\end{tabular}

As an example, suppose $m=6,n=p=3$, then $r=2$. Let $k=3$, $f(x)=x^2-x-1$, $b=(1, 0)^t$, then the new pattern on discovery frame structure
$\mathbb{Z}/6\mathbb{Z} \times \mathbb{Z}/3\mathbb{Z}$ is

\begin{align*}
\begin{array}{ccll}
i(t)& =& i(0)+3t & \mod 6  \\
j(t)& = & j(0)+(c_0(i(0)), c_1(i(0)))b(t) & \mod 3
\end{array}
\end{align*}
where $b(t)$ is the column
vector in $\mathbb{F}_p^{r}$ defined as
\begin{align*}
b(t)=\left\{
\begin{array}{cl}
0 & \mbox{ if } t \equiv 0 \mod p^r  \\
A^{(t \mod p^r)-1}b  & \mbox{ else }
\end{array}
\right.
\end{align*}
where
\begin{align*}
A=
\left(
\begin{array}{cc}
0 & 1 \\
1 & 1
\end{array}
\right) ,
 b=
 \left(
\begin{array}{c}
1\\
0
\end{array}
\right).
\end{align*}
This pattern has period and column period $9$. We show it in the following figure:

\begin{figure}[H]
\centering
\includegraphics[width=\textwidth]{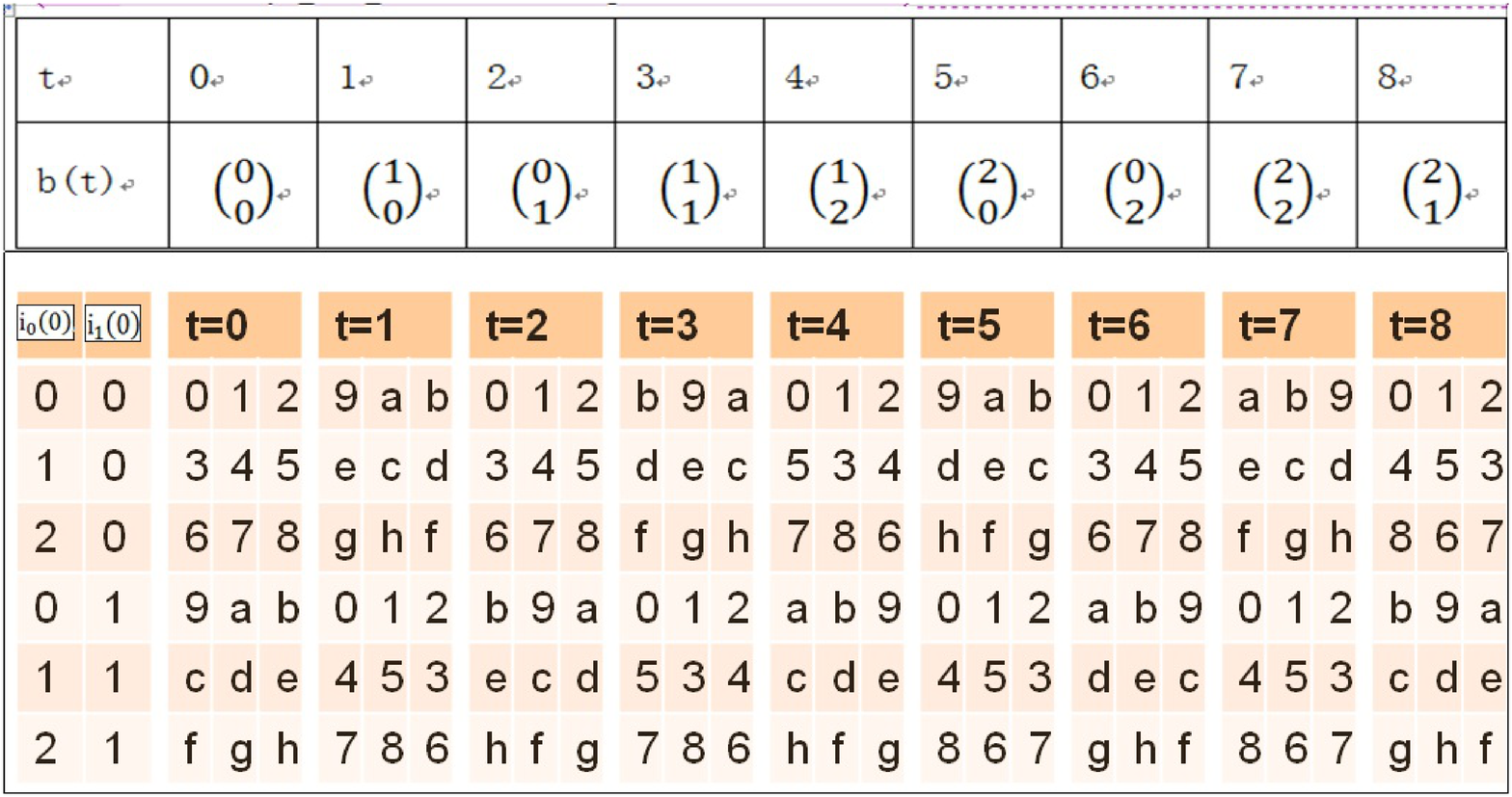}
\caption{A new pattern}
\end{figure}

\section{Simulation result}
We simulate by using the following simulation assumption:

$7 \times 3$ cells, ISD=$500m$, $23$ UEs is uniform dropped per cell, all the pathloss are modeled by O2O NLOS path in Winner$+$
B1 pathloss $PL_{B1}$ with:
\begin{align*}
a. \quad &	h_{BS} = h_{MS} = 1.5m \\
b. \quad &	h_{BS¡¯} = h_{MS¡¯} = 0.8m  \\
c. \quad &	\mbox{offset} = -5 dB.
\end{align*}

The QC hopping pattern, random hopping pattern and the new pattern with
\begin{align*}
& f(x)=x^2+3x+6 \\
& b=(1, 0, \cdots, 0)^t
\end{align*}
on discovery frame structure $\mathbb{Z}/44\mathbb{Z} \times \mathbb{Z}/11\mathbb{Z}$ are simulated.

\begin{figure}[H]
\centering
\includegraphics[width=\textwidth]{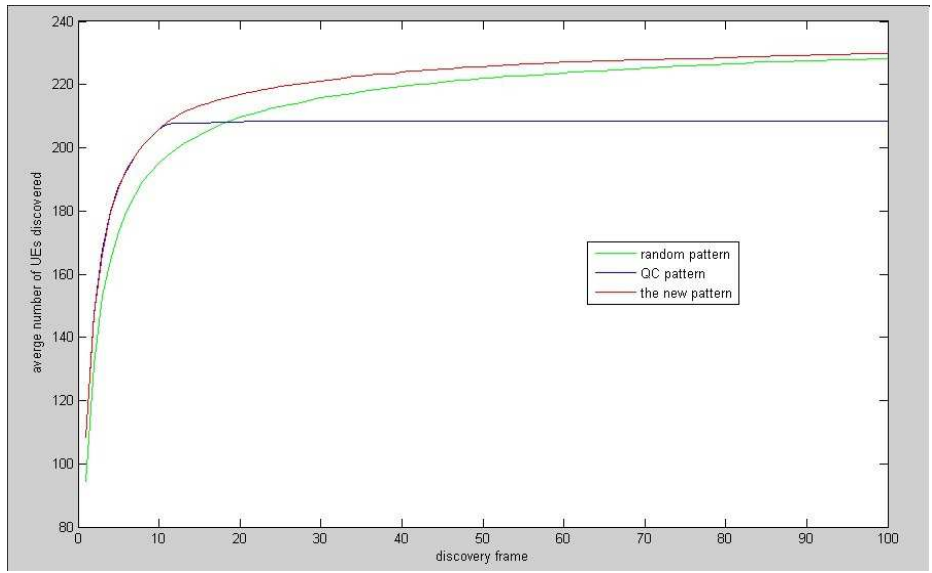}
\caption{Result of simulation}
\end{figure}

We see, when the QC's pattern is used, there are few newly-discovered UEs after $11$th discovery frame, because its
column period is $11$.
This phenomena does not occur when the random pattern or the new pattern is used,
because their column periods are $\infty$ and $121$, respectively.
The discovery speed of the new pattern is faster than the random pattern, because its maximal
 continual collision number is smaller than the random pattern.
The maximal collision ratio measures the fairness, it does not effect the average number of discovered UEs.

\section{Conclusion}
In this paper, We propose three metrics for hopping pattern performance evaluation:
column period, maximal collision ratio, maximal continual collision number. A class of hopping patterns
 is constructed based on the metrics, and through simulation the patterns show better discovery performance.

\end{document}